\documentclass[aps,prd,preprint,superscriptaddress,tightenlines,nofootinbib]{revtex4}
\long\def\symbolfootnote[#1]#2{\begingroup%
\def\thefootnote{\fnsymbol{footnote}}\footnote[#1]{#2}\endgroup}

\usepackage{amsfonts,amsmath,amsthm,amssymb}
\usepackage{mathrsfs}
\usepackage{bm}
\usepackage{color}
\usepackage{axodraw}
\usepackage{epsfig}

\newcommand{\PRE}[1]{{#1}}   
\newcommand{\met} {\not\!\! E_T}
\usepackage{dsfont}
\usepackage{mathcomp}

\newcommand{\postscript}[2]{\setlength{\epsfxsize}{#2\hsize}
   \centerline{\epsfbox{#1}}}
\newcommand{\comment}[1]{}

\newcommand{\el}[1]{\label{#1}}
\newcommand{\er}[1]{\eqref{#1}}

\newcommand{\ci}[1]{}

\newcommand{\ke}{\rangle}
\newcommand{\br}{\langle}

\newcommand{\lsb}{\left[}
\newcommand{\rsb}{\right]}

\newcommand{\nn}{\nonumber \\}
\newcommand{\p}{\partial}

\newcommand{\ba}{\begin{eqnarray}}
\newcommand{\ea}{\end{eqnarray}}
\newcommand{\be}{\begin{equation}}
\newcommand{\ee}{\end{equation}}
\newcommand{\bay}[1]{\left(\begin{array}{#1}}
\newcommand{\eay}{\end{array}\right)}

\def\xt{{\theta}}

\def\CD{{\cal D}}

\def\CR{{\cal R}}

\begin{document}

\preprint{
\hfil
\begin{minipage}[t]{3in}
\begin{flushright}
\vspace*{.4in}
MPP--2011--86\\
LMU-ASC 32/11\\
CERN-PH-TH/2011-180\\
\end{flushright}
\end{minipage}
}


\title{\PRE{\vspace*{0.5in}}
$\bm{Z'}$-gauge Bosons as Harbingers of Low Mass Strings}

\author{{\bf Luis A. Anchordoqui}}
\affiliation{Department of Physics,\\
University of Wisconsin-Milwaukee,
 Milwaukee, WI 53201, USA
\PRE{\vspace*{.05in}}
}

\author{{\bf Ignatios Antoniadis}}\thanks{On leave of absence
from CPHT Ecole Polytechnique, F-91128, Palaiseau Cedex.}
\affiliation{Department of Physics,\\ CERN Theory Division,
CH-1211 Geneva 23, Switzerland
\PRE{\vspace*{.05in}}}

\author{{\bf Haim \nolinebreak Goldberg}}
\affiliation{Department of Physics,\\
Northeastern University, Boston, MA 02115, USA
\PRE{\vspace*{.05in}}
}

\author{{\bf Xing Huang}}
\affiliation{School of Physics and Astronomy, \\
Seoul National University, Seoul 141-747, Korea
\PRE{\vspace*{.05in}}
}

\author{{\bf Dieter L\"ust}}

\affiliation{Max--Planck--Institut f\"ur Physik, \\ 
 Werner--Heisenberg--Institut,
80805 M\"unchen, Germany
\PRE{\vspace*{.05in}}
}

\affiliation{Arnold Sommerfeld Center for Theoretical Physics 
Ludwig-Maximilians-Universit\"at M\"unchen,
80333 M\"unchen, Germany
\PRE{\vspace{.05in}}
}

\author{{\bf Tomasz R. Taylor}}
\affiliation{Department of Physics,\\
  Northeastern University, Boston, MA 02115, USA 
  \PRE{\vspace*{.05in}}
}


\PRE{\vspace*{.15in}}

\begin{abstract}\vskip 2mm
  \noindent
  Massive $Z'$-gauge bosons act as excellent harbingers for string
  compactifications with a low string scale.  In D-brane models they
  are associated to $U(1)$ gauge symmetries that are either anomalous
  in four dimensions or exhibit a hidden higher dimensional anomaly.
  We discuss the possible signals of massive $Z'$-gauge bosons at
  hadron collider machines (Tevatron, LHC) in a minimal D-brane model
  consisting out of four stacks of D-branes.  In this construction,
  there are two massive gauge bosons, which can be naturally
  associated with baryon number $B$ and $B-L$ ($L$ being lepton
  number). Here baryon number is always anomalous in four dimensions,
  whereas the presence of a four-dimensional $B-L$ anomaly depends on
  the $U(1)$-charges of the right handed neutrinos. In case $B-L$ is
  anomaly free, a mass hierarchy between the two associated $Z'$-gauge
  bosons can be explained.  In our phenomenological discussion about
  the possible discovery of massive $Z'$-gauge bosons, we take as a
  benchmark scenario the dijet plus $W$ signal, recently observed by
  the CDF Collaboration at Tevatron. It reveals an excess in the dijet
  mass range $150~{\rm GeV/c}^2$, $4.1\sigma$ beyond SM
  expectations. We show that in the context of low-mass string theory
  this excess can be associated with  the production and decay of a leptophobic $Z'$, a
  singlet partner of $SU(3)$ gluons coupled primarily to baryon
  number.  Even if the CDF signal disappears, as indicated by
  the more recent D0 results, our analysis can still serve as the
  basis for future experimental search for massive $Z'$-gauge bosons
  in low string scale models.  We provide the relevant
  cross sections for the production of $Z'$-gauge bosons in the TeV
  region, leading to predictions that are within reach of the present
  or the next LHC run.

\end{abstract}

\maketitle

\section{Introduction}

Very recently, the CERN Large Hadron Collider (LHC) has fired mankind
into a new era in particle physics. The $SU(3)_C \times SU(2)_L \times
U(1)_Y$ Standard Model (SM) of electroweak and strong interactions was
once again severely tested with a dataset corresponding to an
integrated luminosity of $\sim 4.9~{\rm fb}^{-1}$ of $pp$ collisions
collected at $\sqrt{s} = 7$~TeV. The SM agrees remarkable well with
LHC7 data, but has rather troubling weaknesses and appears to be a
somewhat {\it ad hoc} theory.

It has long been thought that the SM may be a subset of a more
fundamental gauge theory. Several models have been proposed, using the
fundamental principle of gauge invariance as guidepost. A common
thread in most of these proposals is the realization of the SM within
the context of D-brane TeV-scale string
compactifications~\cite{Antoniadis:1998ig}. Such D-brane constructions
extend the SM with several additional $U(1)$
symmetries~\cite{Blumenhagen:2001te}.\footnote{See
  also~\cite{Lebed:2011zg}.}  The basic unit of gauge invariance for
these models is a $U(1)$ field, so that a stack of $N$ identical
D-branes eventually generates a $U(N)$ theory with the associated
$U(N)$ gauge group. (For $N=2$ the gauge group can be $Sp(1) \equiv
SU(2)$ rather than $U(2)$.) Gauge interactions emerge as excitations
of open strings with endpoints attached to the D-branes, whereas
gravitational interactions are described by closed strings that can
propagate in all nine spatial dimensions of string theory (these
comprise parallel dimensions extended along the D-branes and
transverse large extra dimensions).

In this paper we study the main phenomenological aspects of one
particular D-brane model that contains two additional $U(1)$
symmetries, which can be chosen to be mostly baryon number $B$ and
$B-L$, where $L$ is lepton number. This choice is very natural from
the point of view of the SM.  Moreover, with this choice of the two
additional $U(1)$ gauge symmetries, one can obtain a natural mass gap
between the light anomalous $U(1)_B$ gauge boson $Z'$ and the heavier
non-anomalous $U(1)_{B-L}$ gauge boson $Z''$. Our first goal is to
survey the basic features of the gauge theory's prediction regarding
the new mass sector and couplings. These features lead to new
phenomena that can be probed using data from the Tevatron and the LHC.
In particular the theory predicts additional gauge bosons that we will
show are accessible at the hadron colliders.

The layout of the paper is as follows. In Sec.~\ref{SII} we detail
some desirable properties which apply to generic models with multiple
$U(1)$ symmetries. We perform a renormalization group analysis for the
running of the gauge couplings, pointing out that the gauge couplings
of the two group factors $U(1)_a\times SU(N)_a=U(N)_a$ run differently
towards low energies below the string scale. This observation has some
interesting phenomenological consequences. Having so identified the
general properties of the theory, in Sec.~\ref{themodel} we outline
the basic setting of TeV-scale string compactifications and discuss
general aspects of the $U(3)_C \times Sp(1)_L \times U(1)_L \times
U(1)_R$ intersecting D-brane configuration that realize the SM by open
strings. In Secs.~\ref{Tevatron} and \ref{LHC} we
discuss the associated phenomenological aspects of anomalous $U(1)$
gauge bosons related to experimental searches for new physics at the
Tevatron and at the LHC. Finally, in Sec.~\ref{SRegge} we explore
predictions inhereted from properties of the overarching string
theory. Concretely, we study the LHC discovery potential for Regge 
excitations within the D-brane model discussed in this work. Our
conclusions are collected in Sec.~\ref{conclusions}.

\section{Abelian gauge couplings at low energies}
\label{SII}

We begin with the covariant derivative for the $U(1)$  fields in the $1,\, 2,\, 3,\, \dots$ basis in which it is assumed that the kinetic energy terms containing $X_\mu^i$ are canonically normalized
\begin{equation}
{\cal D}_\mu = \partial_\mu - i \sum g'_i \, Q_i  \, X_\mu^i \, .
\label{caldmu1}
\end{equation}
The relations between the $U(1)$ couplings $g'_i$ and any non-abelian counterparts  are left open for now. We carry out an orthogonal transformation of the fields
$X_{\mu}^i = \sum_j {\cal R}_{ij} \, Y_\mu^j$. The covariant derivative becomes
\begin{eqnarray}
{\cal D}_\mu & = & \partial_\mu - i \, \sum_i \sum_j  g'_i \, Q_i  \, {\cal R}_{ij} \, Y_\mu^j \nonumber \\
 & = & \partial_\mu - i \, \sum_j \, \bar g_j \, \bar Q_j \, Y_\mu^j \,,
\end{eqnarray}
where for each $j$
\begin{equation}
\bar g_j \bar Q_j = \sum_i g'_i \, Q_i \, {\cal R}_{ij} \, .
\label{gQ1}
\end{equation}
Next, suppose we are provided with normalization for the hypercharge (taken as $j = 1$)
\begin{equation}
Q_Y = \sum_i c_i \,  Q_i \, ;
\label{normalization}
\end{equation}
hereafter we omit the bars for simplicity.
Rewriting (\ref{gQ1}) for the hypercharge
\begin{equation}
g_Y \,  Q_Y = \sum_i \, g'_i \, Q_i \, {\cal R}_{i1}
\label{gQ2}
\end{equation}
and substituting (\ref{normalization}) into (\ref{gQ2}) we obtain
\begin{equation}
g_Y \, \sum_i \, Q_i \, c_i = \sum_i \, g'_i \, {\cal R}_{i1}  \, Q_i .
\label{3}
\end{equation}

One can think about the charges $Q_{i,p}$ as vectors with the components labeled by particles $p$. Assuming that these vectors are linearly independent, Eq.(\ref{3}) reduces to
\begin{equation}
g_Y \, \, c_i =  \, g'_i \, {\cal R}_{i1}  \,  ,
\end{equation}
or equivalently
\begin{equation}
{\cal R}_{i1} = \frac{g_Y \, c_i}{g'_i} \, .
\label{ocho}
\end{equation}
Orthogonality of the rotation matrix, $\sum_i {\cal R}_{i1}^2 = 1$, implies
\begin{equation}
g_Y^2  \sum_i \left (\frac{c_i}{g'_i} \right)^2 =1 \, .
\end{equation}
Then, the condition
\begin{equation}
P \equiv \frac{1}{g_Y^2} - \sum_i \left(\frac{c_i}{g'_i} \right)^2  =0
\label{perp}
\end{equation}
encodes the orthogonality of the mixing matrix connecting the fields
coupled to the stack charges $Q_1$, $Q_2$, $Q_3,\, \dots$ and the
  fields rotated, so that one of them, $Y$, couples to the hypercharge
  $Q_Y$.

A very important point is that the couplings that are running are those of the $U(1)$ fields; hence the $\beta$ functions receive contributions from fermions and scalars, but not from gauge bosons.
The one loop correction to the various couplings are
\begin{equation} \frac{1}{\alpha_Y(Q)} = \frac{1}{\alpha_Y (M_s)} - \frac{b_Y}{2\pi} \, \ln(Q/M_s) \,, \end{equation} \begin{equation} \frac{1}{\alpha_i(Q)} = \frac{1}{\alpha_i (M_s)} - \frac{b_i}{2\pi} \, \ln(Q/M_s) \,, \label{RGbi} \end{equation} where \begin{equation} b_i = \frac{2}{3} \, {\sum_f} \, Q_{i,f}^2 \, + \frac{1}{3} \, {\sum_s} \, Q_{i,s}^2, \end{equation} 
with $f$ and $s$ indicating contribution from fermion and scalar loops, respectively.

Let us assume that the charges are orthogonal, $\sum_sQ_{i,s}Q_{j,s}=\sum_fQ_{i,f}Q_{j,f}=0$ for $i\neq j$. Then Eq.(\ref{normalization}) implies
\begin{equation}
\sum_sQ_{Y,s}^2=\sum_i c_i^2\sum_sQ_{i,s}^2 \label{ysum}
\end{equation}
and the same thing for fermions, hence
\begin{equation} b_Y = \sum_i c_i^2 b_i\ . \end{equation}  
 On the other, the RG-induced change of $P$ defined in Eq.(\ref{perp}) reads
\begin{eqnarray}
  \Delta P & = & \Delta \left(\frac{1}{\alpha_Y}\right) - \sum_i c_i^2 \, \Delta \left(\frac{1}{\alpha_i}\right) \nonumber \\
             & = & \frac{1}{2\pi} \left(b_Y - \sum_i c_i^2 \, b_i \right) \, \ln(Q/M_s)  \, .
\label{deltap}
\end{eqnarray}
Thus, $P=0$ stays valid to one loop if the charges are orthogonal. An example of orthogonality is seen in the $U(3)_C \times Sp(1)_L \times U(1)$ D-brane model of~\cite{Antoniadis:2004dt, Berenstein:2006pk}, for which the various $U(1)$ assignments  are given in Table~\ref{t1}.
In the 3-stack D-brane models of~\cite{Antoniadis:2000ena}, the charges are linearly independent, but not necessarily orthogonal. If the charges are not orthogonal, the RG equations controlling the running of couplings associated to different charges become coupled. One-loop corrections generate mixed kinetic terms for $U(1)$ gauge fields \cite{delAguila:1988jz}, greatly complicating the analysis.
\begin{table}
\caption{Quantum numbers of  chiral fermions  and  Higgs doublet for  $U(3)_C \times Sp(1)_L \times U(1)$.}
\begin{tabular}{c|ccccc}
\hline
\hline
 Name &~~Representation~~& ~$Q_3$~& ~$Q_1$~ & ~$Q_Y$~ \\
\hline
~~$\bar U_i$~~ & $({\bar 3},1)$ &    $-1$ & $\phantom{-}1$ & $-\frac{2}{3}$ \\[1mm]
~~$\bar D_i$~~ &  $({\bar 3},1)$ &    $-1$ & $-1$ & $\phantom{-}\frac{1}{3}$  \\[1mm]
~~$L_i$~~ & $(1,2)$&    $\phantom{-}0$ & $\phantom{-}1$ & $-\frac{1}{2}$  \\[1mm]
~~$\bar E_i$~~ &  $(1,1)$&  $\phantom{-}0$ & $-2$ &  $\phantom{-}1$  \\[1mm]
~~$Q_i$~~ & $(3,2)$& $\phantom{-}1$ & $\phantom{-}0$ & $\phantom{-}\frac{1}{6}$ \\[1mm]
~~$H$~~ & $(1,2)$ &  $\phantom{-}0$ & $\phantom{-} 1$ & $-\frac{1}{2} $ \\[1mm]
\hline
\hline
\end{tabular}
\label{t1}
\end{table}

Another important element of the RG analysis is that the relation for
$U(N)$ unification, $g_N' = g_N/\sqrt{2N},$ holds only at $M_s$
because the $U(1)$ couplings ($g'_1, g'_2, g'_3$) run differently from
the non-abelian $SU(3)$ ($g_3$) and $SU(2)$ ($g_2$).

\begin{figure}[tbp]
\postscript{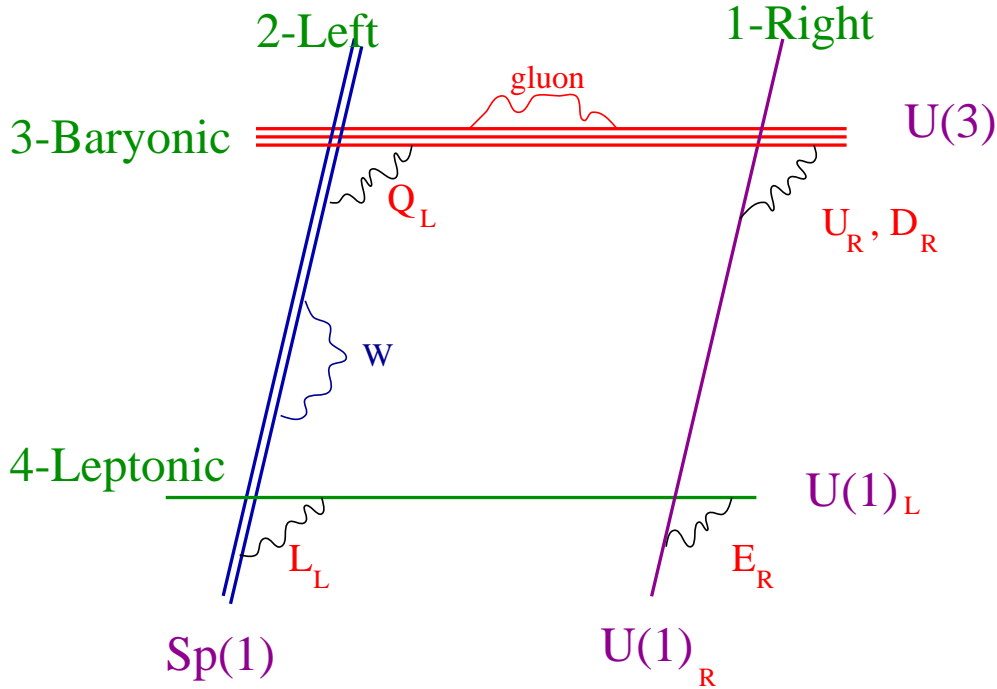}{0.8}
\caption{Pictorial representation of the $U(1)_C \times Sp(1)_L \times U(1)_L \times U(1)_R$ D-brane model.}
\label{4stacks}
\end{figure}

In this paper we are interested in a minimal 4-stack model $U(3)_C
\times Sp(1)_L \times U(1)_L \times U(1)_R$, which has the attractive
property of elevating the two major global symmetries of the SM
(baryon number $B$ and lepton number $L$) to local gauge symmetries.
A schematic representation of the D-brane structure (to be discussed
in detail in Sec.~\ref{themodel}) is shown in Fig.~\ref{4stacks}. The
chiral fermion charges in Table~\ref{t:spectrum} are not orthogonal as
given ($Q_{1L} \cdot Q_{1R} \neq 0,$).  Orthogonality can be completed
by including a right-handed neutrino with charges $Q_3 =0$, $Q_{1L} =
Q_{1R} = \pm 1$, $Q_Y = 0$. We turn now to discuss the string origin
and the compelling properties of this model.

\section{Generalities of $\bm{U(3)_C \times Sp(1)_L \times U(1)_L \times
U(1)_R}$}
\label{themodel}

The generic perturbative spectrum in intersecting D-brane models consists of products of unitary groups $U(N_i)$ associated to stacks of $N_i$ coincident D-branes and matter in bi-fundamental representations. In the presence of orientifolds which are necessary for tadpole cancellation, and thus consistency of the theory, open strings become in general non oriented allowing for orthogonal and symplectic gauge group factors, as well as for symmetric and antisymmetric matter representations.

\begin{table}
\caption{Chiral fermion spectrum of the $U(3)_C \times Sp(1)_L \times U(1)_L \times U(1)_R$ D-brane model.}
\begin{tabular}{c|ccccc}
\hline
\hline
 Name &~~Representation~~& ~$Q_3$~& ~$ Q_{1L}$~ & ~$Q_{1R}$~ & ~~$Q_{Y}$~~ \\
\hline
~~$\bar U_i$~~ & $({\bar 3},1)$ &    $-1$ & $\phantom{-}0$ & $-1$ & $-\frac{2}{3}$  \\[1mm]
~~$\bar D_i$~~ &  $({\bar 3},1)$&    $-1$ & $\phantom{-}0$ & $\phantom{-} 1$ & $\phantom{-}\frac{1}{3}$  \\[1mm]
~~$L_i$~~ & $(1,2)$&    $\phantom{-}0$ &  $\phantom{-}1$ & $\phantom{-}0$ & $-\frac 1 2$  \\[1mm]
~~$\bar E_i$~~ &  $(1,1)$&   $\phantom{-}0$ & $-1$ &  $\phantom{-} 1$ & $\phantom{-} 1$ \\[1mm]
~~$Q_i$~~ & $(3,2)$& $\phantom{-}1$ & $\phantom{-}0 $ & $\phantom{-} 0$ & $\phantom{-}\frac{1}{6}$   \\[1mm]
\hline
\hline
\end{tabular}
\label{t:spectrum}
\end{table}

The minimal embedding of the SM particle spectrum requires at least
three brane stacks~\cite{Antoniadis:2000ena} leading to three distinct
models of the type $U(3)_C\times U(2)_L\times U(1)$ that were
classified in~\cite{Antoniadis:2000ena, Antoniadis:2004dt}. Only one
of them (model C of~\cite{Antoniadis:2004dt}) has Baryon number as
symmetry that guarantees proton stability (in perturbation theory),
and can be used in the framework of TeV strings. Moreover, since $Q_2$
(associated to the $U(1)$ of $U(2)_L$) does not participate in the
hypercharge combination, $U(2)_L$ can be replaced by $Sp(1)_L$ leading
to a model with one extra $U(1)$, the Baryon number, besides
hypercharge~\cite{Berenstein:2006pk}. The quantum numbers of the
chiral SM spectrum are given in Table~\ref{t1}.  Since baryon number
is anomalous, the extra abelian gauge field becomes massive by the
Green-Schwarz (GS) mechanism, behaving at low energies as a $Z'$ with a mass in general lower than the string scale by an order of magnitude corresponding to a loop factor~\cite{Antoniadis:2002cs}. Given the three SM couplings and the hypercharge combination, this model has no free parameter in the coupling of $Z'$ to the SM fields. Moreover, lepton number is not a symmetry creating a problem with large neutrino masses through the Weinberg dimension-five operator $LLHH$ suppressed only by the TeV string scale. We therefore proceed to models with four D-brane stacks.

The SM embedding in four D-brane stacks leads to many more models that have been classified in~\cite{Antoniadis:2002qm, Anastasopoulos:2006da}. In order to make a phenomenologically interesting  choice, we first focus on models where $U(2)_L$ can be reduce to $Sp(1)$. Besides the fact that this reduces the number of extra $U(1)$'s, one avoids the presence of a problematic Peccei-Quinn symmetry, associated in general with the $U(1)$ of $U(2)_L$ under which Higgs doublets are charged~\cite{Antoniadis:2000ena}. We then impose Baryon and Lepton number symmetries that determine completely the model  $U(3)_C \times Sp(1)_L \times U(1)_L \times U(1)_R$, as described in~\cite{Anastasopoulos:2006da} (see subsection 4.2.4). The corresponding fermion quantum numbers are given in Table~\ref{t:spectrum}, while the two extra $U(1)$'s are the Baryon and Lepton number, $B$ and $L$, respectively; they are given by the following combinations:
\ba
B=Q_3/3\quad;\quad L=Q_{1L}\quad;\quad Q_Y={1\over 6}Q_3-{1\over 2}Q_{1L}+{1\over 2}Q_{1R}\, ;
\ea
or equivalently by the inverse relations:
\ba
Q_3=3B\quad;\quad Q_{1L}=L\quad;\quad Q_{1R}=2Q_Y-(B-L)\, .
\label{bb-l}
\ea

Note that with the `canonical' charges of the right-handed neutrino $Q_{1L}=Q_{1R}=-1$, the combination $B-L$ is anomaly free, while for $Q_{1L}=Q_{1R}=+1$, both $B$ and $B-L$ are anomalous. Actually, both choices guarantee orthogonality of the charges discussed in the previous section. As mentioned already, anomalous $U(1)$'s become massive necessarily due to the Green-Schwarz anomaly cancellation, but non anomalous $U(1)$'s can also acquire masses due to effective six-dimensional anomalies associated for instance to sectors preserving $N=2$ supersymmetry~\cite{Antoniadis:2002cs}.\footnote{In fact, also the hypercharge  gauge boson of $U(1)_Y$ can acquire a mass through this mechanism.
In order to keep it massless, certain topological constraints
on the compact space have to be met.} These two-dimensional `bulk' masses become therefore larger than the localized masses associated to four-dimensional anomalies, in the large volume limit of the two extra dimensions. Specifically for D$p$-branes with $(p-3)$-longitudinal compact dimensions the masses of the anomalous and, respectively, the non-anomalous $U(1)$ gauge bosons have the following generic scale
behavior:
\ba
{\rm anomalous}~U(1)_a:~~~M_{Z'}&=&g'_aM_s\, ,\nonumber\\
{\rm non-anomalous}~U(1)_a:~~~M_{Z''}&=&g'_aM_s^3\, V_2\, .
\ea
Here $g'_a$ is the gauge coupling constant associated to the group
$U(1)_a$, given by $g'_a\propto g_s/\sqrt{V_\parallel}$ where $g_s$ is
the string coupling and $V_\parallel$ is the internal D-brane
world-volume along the $(p-3)$ compact extra dimensions, up to an
order one proportionality constant. Moreover, $V_2$ is the internal
two-dimensional volume associated to the effective six-dimensional
anomalies giving mass to the non-anomalous $U(1)_a$. \footnote{It
  should be noted that in spite of the proportionality of the $U(1)_a$
  masses to the string scale, these are not string excitations but
  zero modes.  The proportionality to the string scale appears because the
  mass is generated from anomalies, via an analog of the GS anomaly
  cancellations: either 4 dimensional anomalies, in which case the GS term is
  equivalent to a Stuckelberg mechanism, or from effective 6
  dimensional 
  anomalies, in which case the mass term is extended in two more
  (internal) dimensions. The non-anomalous $U(1)_a$ can also grow a mass
  through a Higgs mechanism. The advantage of the anomaly mechanism versus an
  explicit vev of a scalar field is that the global symmetry survives
  in perturbation theory, which is a desired property for the Baryon
  and Lepton number, protecting proton stability and small neutrino
  masses.}  E.g. for the case of
D5-branes, whose common intersection locus is just 4-dimensional Minkowski-space, $V_\parallel=V_2$ denotes the volume
of the longitudinal, two-dimensional space along the two internal D5-brane directions. 
Since internal volumes are bigger than one in string units to have effective field theory description, the masses of  non-anomalous $U(1)$-gauge bosons are generically larger
than the masses of the anomalous gauge bosons. Since we want to identify the light $Z'$ gauge boson with baryon number, which is always anomalous, a  hierarchy
compared to the second $U(1)$-gauge boson $Z''$ can arise, if we identify  $Z''$ with the anomaly  free combination $B-L$, and take the internal world-volume $V_2$ a bit larger than the
string scale.\footnote{In \cite{Conlon:2008wa} a different (possibly T-dual) scenario with $D7$-branes was investigated. In this case the masses of the anomalous and non-anomalous $U(1)$'s appear to exhibit a dependence on the
entire six-dimensional volume, such that the non-anomalous masses become lighter than the anomalous ones.}
In summary, this model has two free parameters: one coupling and one
angle in the two-dimensional space orthogonal to the hypercharge
defining the direction of the corresponding $Z'$. Tuning the later, it
can become leptophobic, while the former controls the strength of its
interactions to matter. As discussed already, one can distinguish two
cases: (i) when $B$ and $L$ have 4d anomalies, the mass ratio of the
two extra gauge bosons ($Z'$ and $Z''$) is fixed by the ratio of their
gauge couplings, up to order one coefficients; (ii) when $B-L$ is
anomaly free and gets a mass from effective six-dimensional anomalies,
the mass ratio of the leftover anomalous $U(1)$ compared to the
non-anomalous $U(1)$ is suppressed by the two-dimensional volume.

To summarize, we will analyze the phenomenology of two D-brane
constructions with three mutually orthognal $U(1)$ charges, in which
the combination $B-L$ is either anomalous or anomaly free.  In the
next section, we analyze these situations and study the regions of the
parameter space where $Z'$ is leptophobic and can accommodate the
recent Tevatron data.

\section{Leptophobic $\bm{Z'}$ at the Tevatron}
\label{Tevatron}

Taken at face value, the disparity between
CDF~\cite{Aaltonen:2011mk,Punzi} and D0~\cite{Abazov:2011af} results
insinuates a commodious uncertainty as to whether there is an excess
of events in the dijet system invariant mass distribution of the
associated production of a $W$ boson with 2 jets (hereafter $Wjj$
production). The $M_{jj}$ excess showed up in $4.3~{\rm fb}^{-1}$ of
integrated luminosity collected with the CDF detector as a broad bump
between about 120 and 160~GeV~\cite{Aaltonen:2011mk}. The CDF
Collaboration fitted the excess (hundreds of events in the $\ell jj +
\met$ channel) to a Gaussian and estimated its production cross
section times the dijet branching ratio to be $4~{\rm pb}$. This is
roughly 300 times the SM Higgs rate $\sigma (p\bar p
\to WH) \times {\rm BR} (H \to b \bar b)$. For a search window of $120
- 200~{\rm GeV}$, the excess significance above SM background
(including systematics uncertainties) has been reported to be
$3.2\sigma$~\cite{Aaltonen:2011mk}. Recently, CDF has included an
additional $3~{\rm fb}^{-1}$ to their data sample, for a total of
$7.3~{\rm fb}^{-1}$, and the statistical significance has grown to
$\sim 4.8\sigma$ ($\sim 4.1\sigma$ including
systematics)~\cite{Punzi}. More recently, the D0 Collaboration
released an analysis (which closely follows the CDF analysis) of their
$Wjj$ data finding ``no evidence for anomalous resonant dijet
production''~\cite{Abazov:2011af}. Using an integrated luminosity of
$4.3~{\rm fb^{-1}}$ they set a 95\% CL upper limit of $1.9~{\rm pb}$
on a resonant $Wjj$ production cross section.

In a previous work~\cite{Anchordoqui:2011ag} we presented an
explanation of the CDF data by identifying the resonance with a $Z'$
inherent to D-brane TeV-scale string
compactifications~\cite{Antoniadis:1998ig}.  In this section we repeat
our analysis but with two highly significant changes. First, we allow
for the experimental uncertainty by focusing on a wide range ($1.6 -
6.0~{\rm pb}$) of the (pre-cut) $Wjj$ resonant production cross
section. This interpolates between the CDF and D0 results. Second, we
turn our attention to a different D-brane model which has the
attractive property of elevating the two major global symmetries of
the SM (baryon number $B$ and lepton number $L$) to local gauge
symmetries.

Related explanations for the CDF anomaly based on an additional
leptophobic $Z'$ gauge boson have been offered~\cite{Buckley:2011vc}.
Alternative new physics explanations include technicolor, new Higgs
sectors, supersymmetry with and without $R$ parity violation, color
octect production, quirk exchange, and more~\cite
{Eichten:2011sh}. There are also attempts to explain this puzzle
within the context of the SM~\cite{He:2011ss}.

The suppressed coupling to leptons (or more specifically, to electrons
and muons) is required to evade the strong constraints of the Tevatron
$Z'$ searches in the dilepton mode~\cite{Acosta:2005ij} and LEP-II
measurements of $e^+ e^- \to e^+ e^-$ above the
$Z$-pole~\cite{Barate:1999qx}. In complying with the precision
demanded of our phenomenological approach it would be sufficient to
consider a 1\% branching fraction to leptons as consistent with the
experimental bound. This approximation is within a factor of a few of
{\em model independent} published experimental bounds. In addition,
the mixing of the $Z'$ with the SM $Z$ boson should be extremely
small~\cite{Langacker:1991pg,Umeda:1998nq} to be compatible with
precision measurements at the $Z$-pole by the LEP
experiments~\cite{:2005ema}.

All existing dijet-mass searches via direct production at the Tevatron
are limited to $M_{jj} > 200~{\rm GeV}$~\cite{Abe:1993kb} and
therefore cannot constrain the existence of a $Z'$ with $M_{Z'} \simeq
150~{\rm GeV}$. The strongest constraint on a light leptophobic $Z'$
comes from the dijet search by the UA2 Collaboration, which has placed
a 90\% CL upper bound on $\sigma (p\bar p \to Z') \times {\rm BR}(Z'
\to jj)$ in this energy range~\cite{Alitti:1990kw}.  A comprehensive
model independent analysis incorporating Tevatron and UA2 data to
constrain the $Z'$ parameters for predictive purposes at the LHC has
been recently presented~\cite{Hewett:2011nb}.\footnote{Other
  phenomenological restrictions on $Z'$-gauge bosons were recently
  presented in~\cite{Williams:2011qb}.}

In the $U(3)_C \times Sp(1)_L \times U(1)_L \times
U(1)_R$ D-brane model the $Q_3$, $
Q_{1L}$, $Q_{1R}$ content of the hypercharge operator is given by,
\be\el{hyperchargeY} Q_Y = c_1 Q_{1R} + c_3 Q_3 +  c_4 Q_{1L} \, ,\ee
with $c_1 = 1/2,$ $c_3 = 1/6$, and $c_4 =-1/2$.

The covariant derivative  (\ref{caldmu1}) can be re-written as
\be\el{covderi2} \CD_\mu = \p_\mu - i g'_3 \, C_\mu  \,  Q_3   -i  g'_4 \, \tilde B_\mu \,  Q_{1L}  -i g'_1 \, B_\mu \, Q_{1R} \, .\ee
The fields $C_\mu, \tilde B_\mu, B_\mu$ are related
to $Y_\mu, Y_\mu{}'$ and $Y_\mu{}''$ by the rotation matrix,
\begin{equation}
\CR=
\left(
\begin{array}{ccc}
 C_\theta C_\psi  & -C_\phi S_\psi + S_\phi S_\theta C_\psi  & S_\phi
S_\psi +  C_\phi S_\theta C_\psi  \\
 C_\theta S_\psi  & C_\phi C_\psi +  S_\phi S_\theta S_\psi  & - S_\phi
C_\psi + C_\phi S_\theta S_\psi  \\
 - S_\theta  & S_\phi C_\theta  & C_\phi C_\theta
\end{array}
\right) \,,
\end{equation}
with Euler angles $\theta$, $\psi,$ and $\phi$. Equation~(\ref{covderi2}) can be rewritten in terms of $Y_\mu$, $Y'_\mu$, and
$Y''_\mu$ as follows
\begin{eqnarray}
\CD_\mu & = & \partial_\mu -i Y_\mu \left(-S_\xt g'_1 Q_{1R} + C_\theta S_\psi  g'_4  Q_{1L} +  C_\theta C_\psi g'_3 Q_3 \right) \nonumber \\
 & - & i Y'_\mu \left[ C_\theta S_\phi  g'_1 Q_{1R} +\left( C_\phi C_\psi + S_\theta S_\phi S_\psi \right)  g'_4 Q_{1L} +  (C_\psi S_\theta S_\phi - C_\phi S_\psi) g'_3 Q_3 \right] \label{linda} \\
& - & i Y''_\mu \left[ C_\theta C_\phi g'_1 Q_{1R} +  \left(-C_\psi S_\phi + C_\phi S_\theta S_\psi \right)  g'_4  Q_{1L} + \left( C_\phi C_\psi S_\theta + S_\phi S_\psi\right) g'_3 Q_3 \right]   \, .  \nonumber
\end{eqnarray}
Now, by demanding that $Y_\mu$ has the
hypercharge $Q_Y$ given in Eq.~\er{hyperchargeY}  we  fix the first column of the rotation matrix $\CR$
\begin{equation}
\bay{c} C_\mu \\ \tilde B_\mu \\ B_\mu
\eay = \left(
\begin{array}{lr}
  Y_\mu \, c_3g_Y /g'_3& \dots \\
  Y_\mu \, c_4 g_Y/g'_4 & \dots\\
   Y_\mu \, c_1g_Y/g'_1 & \dots
\end{array}
\right) \, ,
\end{equation}
and we determine the value of the two associated Euler angles
\begin{equation}
\theta = {\rm -arcsin} [c_1 g_Y/g'_1]
\label{theta}
\end{equation}
and
\begin{equation}
\psi = {\rm arcsin}  [c_4 g_Y/ (g'_4 \, C_\theta)] \, .
\label{psi}
\end{equation}
The couplings $g'_1$ and $g'_4$ are related through the orthogonality condition (\ref{perp}),
\begin{equation}
 \left(\frac{c_4}{ g' _4} \right)^2  = \frac{1}{g_Y^2} - \left(\frac{c_3}{g'_3} \right)^2  - \left(\frac{c_1}{g'_1}\right)^2  \, ,
\end{equation}
with $g'_3$ fixed by the relation $g_3 (M_s) = \sqrt{6} \, g'_3 (M_s)$. In what follows, we take $M_s = 5~{\rm TeV}$ as a reference point for running down to 150~GeV the $g'_3$ coupling using (\ref{RGbi}), that is {\em ignoring mass threshold effects of stringy states}. This yields $g'_3 = 0.383$. We have checked that the running of the $g'_3$ coupling does not change significantly within the LHC range, i.e., $3~{\rm TeV} < M_s < 10~{\rm TeV}.$

The phenomenological analysis thus far has been formulated in terms
of the  mass-diagonal basis set of gauge fields $(Y,Y',Y'')$. As a
result of the electroweak phase transition, the coupling of this set
with the Higgses will induce mixing, resulting in a new mass-diagonal basis set
$(Z,Z',Z'')$. It will suffice to analyze only the $2\times 2$ system
$(Y,Y')$ to see that the effects of this mixing are totally
negligible. We consider simplified zeroth and first order (mass)$^2$
matrices
\begin{equation}
 (M^2)^{(0)}\ =\ \left (\begin{array}{cc} 0 & 0 \\ 0 & M'^2 \end{array} \right) \quad
(M^2)^{(1)}\ = \ \left (\begin{array}{cc} \overline{M}_Z^2 & \epsilon \\ \epsilon & m'^2 \end{array} \right)
\end{equation}
where $M'$ is the mass of the $Y'$ gauge field, $\overline{M}_Z = \sqrt{(g_2^2 v^2 + g_Y^2 v^2)/2}$ is the usual tree level formula for the mass  of the $Z$ particle in the electroweak theory (before mixing), $g_2$ is the electroweak coupling constant, $v$ is the vacuum expectation value of the Higgs field, and $\epsilon, m'^2$ are of ${\cal O}(\overline{M}_Z^2)$.

Standard Rayleigh-Schrodinger perturbation theory then provides the (mass)$^2$ (to second order in
$\overline{M}_Z^2$) and
wave functions (to first order)  of the mass-diagonal eigenfields $(Z,Z')$ corresponding to $(Y,Y')$.
\begin{equation}
M_Z^2 = \overline{M}_Z^2 - \left(\frac{\epsilon^2}{M'^2}\right) \,, \quad \quad
M_{Z'}^2 = M'^2 + m'^2 + \left(\frac{\epsilon^2}{M'^2}\right) \,,
\label{seag1}
\end{equation}
and
\begin{equation}
Z = Y - \left( \frac{\epsilon}{M'^2} \right)  \ \ Y' \,, \quad \quad
Z' = Y' +  \left(\frac{\epsilon}{M'^2} \right)\ \ Y  \, .
\label{seag2}
\end{equation}
From Eqs.~(\ref{seag1}) and (\ref{seag2}) the shift in the mass of the $Z$ is given by $\delta M_Z^2 = (\epsilon/M')^2,$ so that
$\epsilon = M' \sqrt{2 M_Z \delta M_Z}.$ The admixture of $Y$ in the mass-diagonal field $Z'$ is
\begin{equation}
 \theta = \frac{\epsilon}{M'^2}  = \frac{M_Z}{M'} \;\;\sqrt{\frac{2 \delta M_Z} {M_Z}} \simeq 0.004 \, ,
\end{equation}
where we have taken $\delta M_Z = 0.0021~{\rm
  GeV}$~\cite{Nakamura:2010zzi}.  Interference effects which are
proportional to $\theta$ are present in processes with fermions
(e.g. Drell-Yan).  However, these vanish at the peak of the
resonance. Because of the smallness of $\theta$, modifications of SM
partial decay rates of the $Z$ are negligible. (See
e.g.~\cite{Langacker:1991pg}, for an analysis of such effects.)
Remaining effects are order $\theta^2\simeq 1.6\times 10^{-5}$, and
therefore all further discussion will be, with negligible error, in
terms of $Z'$.  By the same token, the admixture of $Y'$ in the
eigenfield $Z$ is negligible, so that the discussion henceforth will
reflect $Z \simeq Y$ and $\overline{M}_Z^2 \simeq M_Z^2$.

The third Euler angle $\phi$ and the coupling $g'_1$ are determined by
requiring sufficient suppression ($\alt 1\%$) to leptons, a (pre-cut) 
production rate $1.6 \alt \sigma (p\bar p \to WZ') \times {\rm BR} (Z'\to jj)
\alt 6.0~{\rm pb}$ at $\sqrt{s} = 1.96~{\rm TeV}$, and compatibility
with the 90\%CL upper limit reported by the UA2 Collaboration on
$\sigma (p\bar p \to Z') \times {\rm BR} (Z' \to jj)$ at $\sqrt{s} =
630~{\rm GeV}$~\cite{Alitti:1990kw}. 

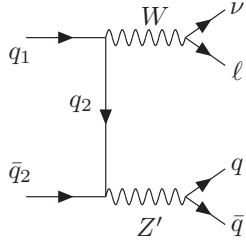
\begin{figure}[t]
\vspace*{1.0cm}
\[
\phantom{XXXXXXX}
\vcenter{
\hbox{
  \begin{picture}(0,0)(0,0)
\SetScale{1.5}
  \SetWidth{.3}
\ArrowLine(-45,20)(-25,20)
\Photon(-25,20)(-5,20){2}{6}
\ArrowLine(-45,-20)(-25,-20)
\Photon(-5,-20)(-25,-20){2}{6}
\ArrowLine(-25,20)(-25,-20)
\ArrowLine(-5,-20)(5,-13)
\ArrowLine(-5,-20)(5,-27)
\ArrowLine(-5,20)(5,27)
\ArrowLine(-5,20)(5,13)
\Text(-70,20)[cb]{{\footnotesize $q_1$}}
\Text(-70,-25)[cb]{{\footnotesize $\bar q_2$}}
\Text(-21,-45)[cb]{{\footnotesize $Z'$}}
\Text(-19,35)[cb]{{\footnotesize $W$}}
\Text(12,-22)[cb]{{\footnotesize $q$}}
\Text(12,-45)[cb]{{\footnotesize $\bar q$}}
\Text(12,39)[cb]{{\footnotesize $\nu$}}
\Text(-47,2)[cb]{{\footnotesize $q_2$}}
\Text(12,14)[cb]{{\footnotesize $\ell$}}
\end{picture}}
}
\]
\vspace*{.6cm}
\caption[]{Feynman diagram for $q\bar q \to WZ' \to \nu \ell jj$.}
\label{feynman}
\end{figure}

The $f \bar f Z'$ Lagrangian is of the form
\begin{eqnarray}
{\cal L} & = & \frac{1}{2}   \sqrt{g_Y^2 + g_2^2} \ \sum_f \bigg(\epsilon_{f_L} \bar \psi_{f_L} \gamma^\mu \psi_{f_L} +   \epsilon_{f_R} \bar \psi_{f_R} \gamma^\mu \psi_{f_R} \bigg) \, Z'_\mu \, \nonumber \\
& = & \sum_f \bigg((g_{Y'}Q_{Y'})_{f_L} \, \bar \psi_{f_L} \gamma^\mu \psi_{f_L} +  (g_{Y'}Q_{Y'})_{f_R} \bar \psi_{f_R} \gamma^\mu \psi_{f_R} \bigg) \, Z'_\mu \,
\label{lagrangian}
\end{eqnarray}
where each $\psi_{f_{L \, (R)}}$ is a fermion field  with the corresponding $\gamma^\mu$  matrices of the Dirac algebra, and $\epsilon_{f_L,f_R} = v_q \pm a_q$, with $v_q$ and $a_q$ the vector and axial couplings respectively.  The (pre-cut) $Wjj$ production rate  at the Tevatron $\sqrt{s} = 1.96~{\rm pb}$, for arbitrary couplings and $M_{Z'} \simeq 150~{\rm GeV}$, is found to be~\cite{Hewett:2011nb}
\begin{equation}
\sigma (p\bar p \to WZ') \times {\rm BR} (Z' \to jj)  \simeq  \left[0.719 \left(\epsilon_{u_L}^2 + \epsilon_{d_L}^2\right) + 5.083 \, \epsilon_{u_L}  \, \epsilon_{d_L} \right] \times \Gamma (\phi, g'_1)_{Z'\to q \bar q}~{\rm pb}  \, ,
\label{wjj-production}
\end{equation}
where $\Gamma (\phi, g'_1)_{Z'\to q \bar q}$ is the hadronic branching fraction. The presence of a $W$ in the process shown in Fig.~\ref{feynman} restricts the contribution of the quarks to be purely left-handed.
The dijet  production rate at the UA2 $\sqrt{s} = 630~{\rm GeV}$ can be parametrized as follows~\cite{Hewett:2011nb}
\begin{equation}
\sigma (p\bar p \to Z' ) \times {\rm BR} (Z'\to jj) \simeq \tfrac{1}{2} \big[ 773  ( \epsilon_{u_L}^2 + \epsilon_{u_R} ^2) + 138  ( \epsilon_{d_L}^2 + \epsilon_{d_R} ^2)\big] \times \Gamma(\phi,g'_1)_{Z' \to q  \bar q}~{\rm pb} \, .
\label{zprime-production}
\end{equation}
(Our numerical calculation using  CTEQ6~\cite{Pumplin:2002vw} agrees within 5\% with the result of~\cite{Hewett:2011nb}.)
The dilepton production rate at UA2 energies is given by
\begin{equation}
\sigma (p\bar p \to Z' ) \times {\rm BR} (Z'\to \ell \bar \ell) \simeq \tfrac{1}{2} \big[ 773  ( \epsilon_{u_L}^2 + \epsilon_{u_R} ^2) + 138  ( \epsilon_{d_L}^2 + \epsilon_{d_R} ^2)\big] \times \Gamma (\phi, g'_1)_{Z' \to \ell  \bar \ell}~{\rm pb} \, ,
\label{zprimeleptons}
\end{equation}
where $\Gamma(\phi, g'_1)_{Z' \to \ell  \bar \ell}$ is the leptonic branching fraction. From (\ref{linda}) and (\ref{lagrangian}) we obtain the explicit form of the chiral couplings in terms of $\phi$ and $g'_1$
\begin{eqnarray}
\epsilon_{u_L} = \epsilon_{d_L} & = & \frac{2}{\sqrt{g_Y^2 + g_2^2}} \, (C_\psi S_\theta S_\psi - C_\phi S_\psi) g'_3 \,, \nonumber \\
\epsilon_{u_R} & = &- \frac{2}{\sqrt{g_Y^2 + g_2^2}} \, [C_\theta S_\phi g'_1 + (C_\psi S_\theta S_\psi - C_\phi S_\psi) g'_3] \,, \label{couplingphig}\\
\epsilon_{d_R} & = & \frac{2}{\sqrt{g_Y^2 + g_2^2}} \, [C_\theta S_\phi g'_1 - (C_\psi S_\theta S_\psi - C_\phi S_\psi) g'_3] \, . \nonumber
\end{eqnarray}
Using (\ref{wjj-production}), (\ref{zprime-production}),
(\ref{zprimeleptons}), and (\ref{couplingphig}) the ratio of branching ratios of
electrons to quarks is minimized within the $\phi-g'_1$ parameter
space, subject to sufficient $Wjj$ production and saturation of the
90\%CL upper limit. For a (pre-cut) $Wjj$ production varying between
$1.6 - 6.0~{\rm pb}$, one possible allowed region of the $\phi - g'_1$
parameter space is found to be $-1.16 \alt \phi \alt 2.12$ and $0.20
\alt g'_1 \alt 0.27$.

\subsection{Anomalous $\bm{B-L}$}

Let us first consider  a reference point of the $\phi - g'_1$ parameter
space consistent with the recent D0 limit~\cite{Abazov:2011af}. For
$\phi = -1.16$ and $g'_1 = 0.27$, corresponding to a suppression
$\Gamma_{Z' \to e^+ e^-}/\Gamma_{Z'\to q \bar q} \sim 1\%$, we obtain
$\sigma (p\bar p \to WZ') \times {\rm BR} (Z'\to jj) \simeq 1.6~{\rm
  pb}$ at $\sqrt{s} = 1.96~{\rm TeV}$. From Eqs.~(\ref{theta}) and
(\ref{psi}), this also corresponds to $\theta = -0.722$, $\psi =
-1.37$.\footnote{The UA2 data has a dijet mass resolution $\Delta
  M_{jj}/M_{jj} \sim 10\%$~\cite{Alitti:1990kw}. Therefore, at 150 GeV
  the dijet mass resolution is about 15~GeV. This is much larger than
  the resonance width, which is calculated to be $\Gamma(Z' \to f \bar
  f) \simeq 5~{\rm GeV}$~\cite{Barger:1996kr}.} All the couplings of
the $Y'$ (or equivalently $Z'$) gauge boson are now detemined and
contained in Eq.~(\ref{linda}). Numerical values are given in
Table~\ref{case1} under the heading of $g_{Y'} Q_{Y'}$.

\begin{figure}[tbp]  
\postscript{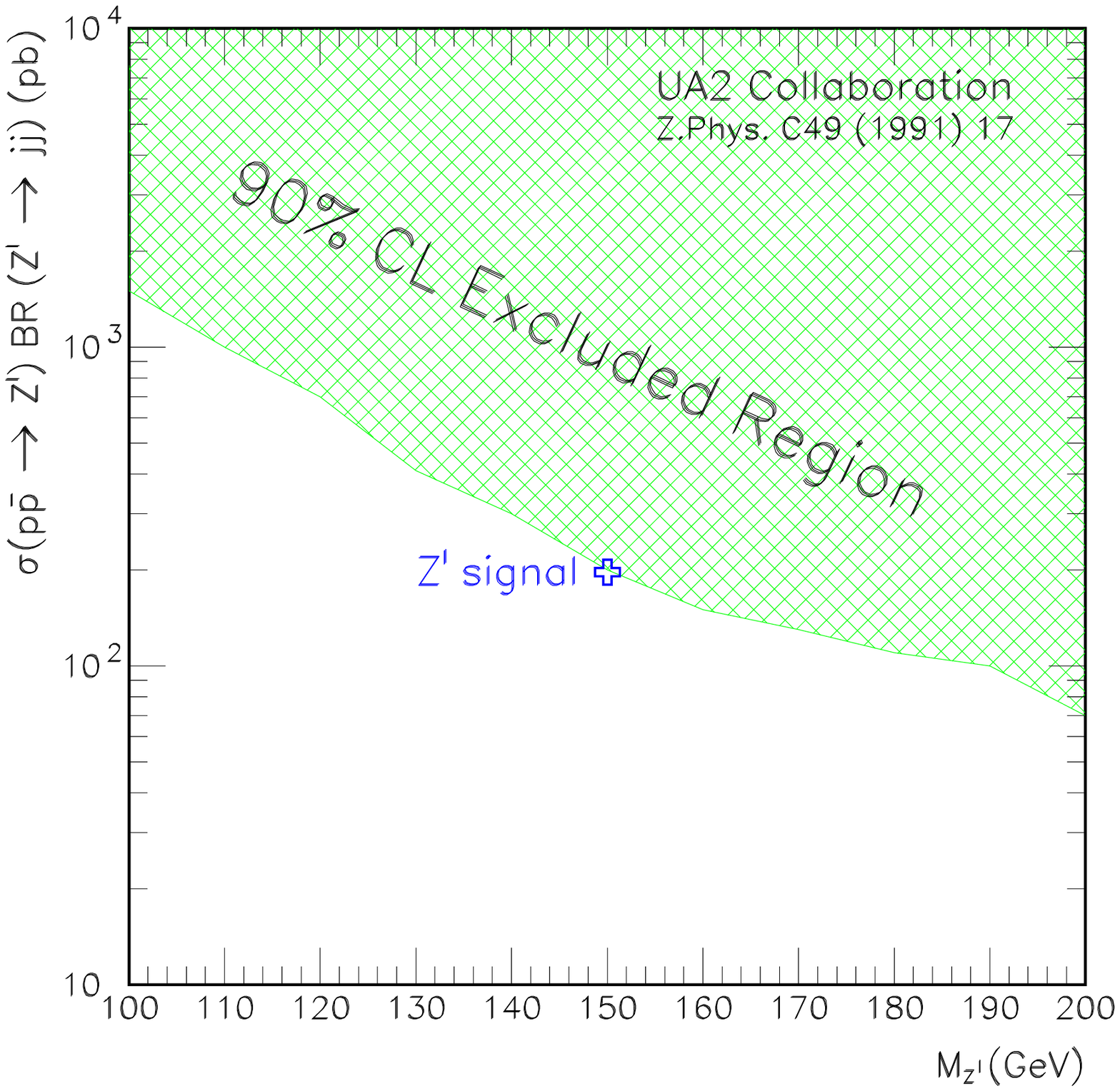}{0.99} 
 \caption{Comparison of the total cross section for the production of $p\bar p \to Z' \to jj$ at $\sqrt{s} = 630~{\rm GeV}$ and the UA2 90\% CL upper limit on the production of a gauge boson decaying into two jets~\cite{Alitti:1990kw}. We have taken $\phi = -1.16$ and $g'_1 = 0.27$.}
\label{UA2}
\end{figure}

In Fig.~\ref{UA2} we show a comparison of $\sigma(p\bar p \to Z')
\times {\rm BR} (Z'\to jj)$ at $\sqrt{s} = 630~{\rm GeV}$ and the UA2
90\% CL upper limit on the production of a gauge boson decaying into
two jets. One can see that for our fiducial values, $\phi = -1.16$ and
$g'_1 = 0.27$, the single $Z' \to jj$ production cross section
saturates the UA2 90\% CL upper limit.

The Tevatron rate for the associated production channels~\cite{Hewett:2011nb}
\begin{equation}
\sigma(p \bar p  \to  ZZ') \times {\rm BR} (Z' \to jj) \simeq \tfrac{1}{4} \big[ 381.5 \epsilon_{u_L}^2 + 221 \epsilon_{u_R}^2 + 1323 \epsilon_{d_L}^2 + 44.1 \epsilon_{d_R}^2 \big] \times \Gamma_{Z' \to q  \bar q}~{\rm fb} \,
\label{ZZprime}
\end{equation}
and
\begin{equation}
\sigma (p \bar p  \to  \gamma Z') \times {\rm BR} (Z' \to jj) \simeq \tfrac{1}{2} \left[ 767 (\epsilon_{u_L}^2 + \epsilon_{u_R}^2 )+ 72.7 (\epsilon_{d_L}^2 +  \epsilon_{d_R}^2)\right] \times \Gamma_{Z' \to q  \bar q}~{\rm fb}
\end{equation}
is always substantially smaller.
(In (\ref{ZZprime})  the SM leptonic branching fractions have been included to ease comparison with the experiment.) It is straightforward to see that these processes should not yet have been  observed at the Tevatron.

\begin{table}
\caption{Chiral couplings of $Y'$ and $Y''$ gauge bosons for $\phi = -1.16$ and $g'_1 = 0.27$.}
\begin{tabular}{c|cc}
\hline
\hline
 Name &~~ $g_{Y'}Q_{Y'}$~ &~~$g_{Y''} Q_{Y''}$\\
\hline
~~$\bar U_i$~~ & $-0.013$ & $-0.411$ \\[1mm]
~~$\bar D_i$~~ &  $-0.386$ &  $-0.251$ \\[1mm]
~~$L_i$~~ &  $-0.125$ & $-0.125$\\[1mm]
~~$\bar E_i$~~ & $-0.061$ & $-0.027$\\[1mm]
~~$Q_i$~~ &  $\phantom{-}0.199$ & $ \phantom{-} 0.331$ \\[1mm]
\hline
\hline
\end{tabular}
\label{case1}
\end{table}

The second strong constraint on the model derives from the mixing of the $Z$ and the $Y'$ through their coupling to the two Higgs doublets $H_1$ and $H_2$. The criteria we adopt here to define the Higgs charges is to make the Yukawa couplings ($H_u \bar u q$, $H_d \bar d q$, $H_d \bar e \ell,$ $H_\nu \bar \nu \ell$) invariant under all three $U(1)$'s. From Table~\ref{t:spectrum}, $\bar u q$ has the charges ($0, 0, -1)$ and $\bar d q$ has ($0, 0, 1$). So the Higgs $H_u$ has $Q_3 = Q_{1L} = 0$, $Q_{1R} = 1$, $Q_Y = 1/2$,  whereas $H_d$ has opposite charges
$Q_3 = Q_{1L} = 0$, $Q_{1R} = -1$, $Q_Y = -1/2$.

Let us consider first the case in which the right-handed neutrino has the following $U(1)$ charges $(0,1,1)$.  As explained before,
$B-L$ is then anomalous, and there is no hierarchy among the masses of $Z'$ and $Z''$. For simplicity we can assume that $H_u\equiv H_1$ and $H_d=H_1^*$, 
with $\br H_1 \ke = (^0_{v_1})$.
For  the second Higgs field $H_\nu\equiv H_2$ the charges are $Q_3 = 0$, $Q_{1L} = -2$, $Q_{1R} = -1$, $Q_Y = 1/2$.\footnote{Note that $H_2$ cannot 
correspond to an elementary open string excitation, since it has $Q_{1L} = -2$. One possibility is to regard $H_2$ as a composite scalar field, built from
two elementary open string scalars, a SM singlet $\phi$ and another Higgs doublet $H_2'$, $H_2\sim\phi H_2'$, with the following $U(1)$-charges: $\phi: \, (0,-1,-1)$ and $H_2': \, (0,-1,0)$.
In case $H_2$ is a composite scalar field so that the corresponding Yukawa coupling arises from a dimension-5 effective operator, one expects that its vacuum expectation value is somewhat suppressed compared to the vev of $H_1$, i.e. $\tan \delta \equiv v_1/v_2>1$.}
Here, $\br H_2 \ke = (^0_{v_2})$, $v = \sqrt{v_1^2 + v_2^2} = 174~{\rm GeV}$, and $\tan \delta \equiv v_1/v_2$. The last
two terms in the covariant derivative
\begin{equation}
\CD_\mu = \p_\mu - i \frac{1}{\sqrt{g_2^2 + g_Y^2}} Z_\mu (g_2^2 T^3 - g_Y^2 Q_Y) -i g_{Y'} Y_\mu{}' Q_{Y'} -i g_{Y''}
Y_\mu{}'' Q_{Y''} ,
\end{equation}
are conveniently written as
\begin{equation}
-i \frac {x_{H_i}}{ v_i} \overline M_Z Y_\mu{}' - i \frac{y_{H_i}} {v_i} \overline M_Z Y_\mu{}'' 
\end{equation}
for each Higgs $H_i$, with $T^3 = \sigma^3/2$, where for the two Higgs doublets
\begin{equation}
x_{H_1} = 1.9 \ \sqrt{{g'_1}^2 -0.032} \ S_\phi \,, \end{equation}
\begin{equation}
   x_{H_2} = - x_{H_1} -1.9 \left[0.054g_1'{}^2 \sqrt{\frac{ 1}{ ({g'_1}^2-0.032)({g'_1}^2-0.033)}}
      \  C_\phi + \frac{0.064 \ S_\phi}{
      \sqrt{{g'_1}^2 -0.032} } \right] \,,
\end{equation}
\begin{equation}
y_{H_1} =  1.9 \ \sqrt{{g'_1}^2 -0.032}
 \ C_\phi \,,
\end{equation}
and
\begin{equation}
y_{H_2} =  - y_{H_1} - 1.9 \left[0.054g_1'{}^2 \sqrt{\frac{ 1}{ ({g'_1}^2-0.032)({g'_1}^2-0.033)}} \
        S_\phi + \frac{0.064 \ C_\phi}{
      \sqrt{{g'_1}^2 - 0.032} } \right] \, .
\end{equation}
For our fiducial values of $\phi$ and $g'_1$ we obtain $x_{H_1} =  - 0.351$, $x_{H_2} = 0.822$,  $y_{H_1}=  0.151$, and $y_{H_2} = -0.556$.

The Higgs field  kinetic term together with the Green-Schwarz mass terms  ($-\frac{1}{2} M'^2 Y'_\mu Y'^\mu - \frac{1}{2} M''^2 Y''_\mu Y''^\mu$, see Appendix) yield the following mass square matrix
$$ \bay{ccc} \overline M_Z^2 & \overline M_Z^2 (x_{H_1} C_\delta^2 + x_{H_2} S_\delta^2) & \overline M_Z^2 (y_{H_1}  C_\delta^2 + y_{H_2} S_\delta^2) \\ \overline M_Z^2 (x_{H_1} C_\delta^2 + x_{H_2} S_\delta^2) &
\overline M_Z^2 (C_\delta^2 x_{H_1}^2 + S_\delta^2 x_{H_2}^2) + M'^2 & \overline M_Z^2
(C_\delta^2 x_{H_1} y_{H_1} + S_\delta^2  x_{H_2} y_{H_2})  \\
\overline M_Z^2 (y_{H_1} C_\delta^2 + y_{H_2} S_\delta^2) & \overline M_Z^2 (C_\delta^2 x_{H_1} y_{H_1} + S_\delta^2 x_{H_2} y_{H_2}) & \overline M_Z^2 (y_{H_1}^2 C_\delta^2 + y_{H_2}^2 S_\delta^2) + M''^2\eay \, .$$
The free parameters are $\tan \delta$, $M_{Z'},$ and $M_{Z''}$ which will be fixed  by requiring  the shift of the $Z$ mass to lie within 1 standard deviation of the experimental value and $M_{Z'} = 150 \pm 5~{\rm GeV}$. We are also minimizing $M_{Z''}$  to ascertain whether it can be detected at existing colliders. This leads to $\tan \delta = 0.65$ and $M_{Z''} \simeq M'' \geq 0.90~{\rm TeV}$.

\begin{table}
\caption{Chiral couplings of $Y'$ and $Y''$ gauge bosons for $\phi = 2.12$ and $g'_1 = 0.26$.}
\begin{tabular}{c|cc}
\hline
\hline
 Name &~~ $g_{Y'}Q_{Y'}$~ &~~$g_{Y''} Q_{Y''}$\\
\hline
~~$\bar U_i$~~ & $\phantom{-}0.088$ & $\phantom{-} 0.395$ \\[1mm]
~~$\bar D_i$~~ &  $\phantom{-} 0.410 $ &  $\phantom{-} 0.197$ \\[1mm]
~~$L_i$~~ & $\phantom{-}0.116$ & $\phantom{-} 0.116$\\[1mm]
~~$\bar E_i$~~ & $\phantom{-} 0.045$ & $\phantom{-} 0.034$\\[1mm]
~~$Q_i$~~ &  $-0.249$ & $ -0.296$ \\[1mm]
\hline
\hline
\end{tabular}
\label{case2}
\end{table}

Next, we scan the parameter space to obtain a larger $Wjj$ production
cross section at the Tevatron. For $\phi = 2.12$ and $g'_1 = 0.26$,
corresponding to a suppression $\Gamma_{Z' \to e^+ e^-}/\Gamma_{Z'\to
  q \bar q} \sim 0.6\%$, $\theta = -0.76$, and $\psi = -1.36$, one
obtains $\sigma (p \bar p \to WZ') \times {\rm BR} (Z' \to jj) =
2.9~{\rm pb}.$ The associated $g_{Y'}Q_{Y'}$ and $g_{Y''} Q_{Y''}$
couplings are given in Table~\ref{case2}. It is straightforward to see
that for $x_{H_1} = 0.303$, $x_{H_2} = -0.740$, $y_{H_1} = -0.187$,
and $y_{H_2} = 0.690$ the shift of the $Z$ mass would lie within 1
standard deviation of the experimental value if $\tan \delta \simeq 0.64$,
$M_{Z'} = 150~{\rm GeV}$, and $M_{Z''} \geq 0.92~{\rm TeV}$.

\subsection{Non-anomalous $\bm{B-L}$}

We now turn to discuss the alternative framework where the the $U(1)$
right-handed neutrino charges are $Q_3 = 0,$ $Q_{1L}= Q_{1R} = -1$,
which means that $B-L$ is anomaly free. Therefore this case is
somewhat preferred compared to the previous case, since there can be a
natural hierarchy among $Z'$ and $Z''$.  In fact, as we will show below, the mixing angle $\phi$ is small and therefore $Z'$ and $Z''$ become essentially  $B$ and $B-L$, respectively.
In such a case, two
`supersymmetric' Higgses $H_u\equiv H_\nu$ and $H_d$ with charges $Q_3
= Q_{1L} = 0$, $Q_{1R} = 1$, $Q_Y = 1/2$ and $Q_3 = Q_{1L} = 0$,
$Q_{1R} = -1$, $Q_Y = -1/2$ would be sufficient to give masses to all
the chiral fermions.  Here, $\br H_u \ke = (^0_{v_u})$, $\br H_d \ke
= (^{v_d}_0),$ $v = \sqrt{v_u^2 + v_d^2} = 174~{\rm GeV}$, and
$\tan \beta \equiv v_u/v_d$. For this particular selection of $U(1)$
charges $x_{H_u} = - x_{H_d} = x_{H_1}$ and $y_{H_u} = - y_{H_d} =
y_{H_1}$. Therefore, it is straightforward to see that the
corresponding mass square matrix for the $Z-Z'$ mixing,
$$ \bay{ccc} \overline M_Z^2 & \overline M_Z^2 (x_{H_u} C_\beta^2 - x_{H_d} S_\beta^2) & \overline M_Z^2 (y_{H_u}  C_\beta^2 - y_{H_d} S_\beta^2) \\ \overline M_Z^2 (x_{H_u} C_\beta^2 - x_{H_d} S_\beta^2) &
\overline M_Z^2 (C_\beta^2 x_{H_u}^2 + S_\beta^2 x_{H_d}^2) + M'^2 & \overline M_Z^2
(C_\beta^2 x_{H_u} y_{H_u} + S_\beta^2  x_{H_d} y_{H_d})  \\
\overline M_Z^2 (y_{H_u} C_\beta^2 - y_{H_d} S_\beta^2) & \overline M_Z^2 (C_\beta^2 x_{H_u} y_{H_u} + S_\beta^2 x_{H_d} y_{H_d}) & \overline M_Z^2 (y_{H_u}^2 C_\beta^2 + y_{H_d}^2 S_\beta^2) + M''^2\eay \,, $$ does not impose any constraint on the $\tan \beta$ parameter. We then use the two degrees of freedom of the model $(g'_1, \phi)$ to demand the shift of the $Z$ mass to lie within 1 standard deviation of the experimental value and leptophobia. Taking $M_{Z'} = 150~{\rm GeV}$, with $g'_1 = 0.20$, $\phi = 0.0028$, and $M_{Z''} = 5~{\rm TeV}$, we find that $\Gamma_{Z'' \to e^+ e^-}/\Gamma_{Z''\to q \bar q} \simeq 1\%$. 

Recall that for
this particular $U(1)$ charge selection of the right-handed neutrino
the combination $B-L$ is anomaly free. Therefore, the mass ratio of
the anomalous and the non-anomalous $U(1)$ can be ascribed to a
suppression induced by the large two-dimensional volume. The
$g_{Y'}Q_{Y'}$ and $g_{Y''} Q_{Y''}$ couplings to the chiral fields
are fixed and given in Table~\ref{case3}. The $Z'$ couplings to quarks
leads to a large (pre-cut) $Wjj$ production ($\simeq 6~{\rm pb}$) at
the Tevatron, and at $\sqrt{s} = 630~{\rm GeV}$, a direct (pre-cut)
$Z' \to jj$ production ($\simeq 700~{\rm pb}$) in the region excluded
by UA2 data. However, it is worthwhile to point out that the UA2
Collaboration performed their analysis in the early days of QCD jet
studies. Their upper bound depends crucially on the quality of the
Monte Carlo and detector simulation which are primitive by today's
standard. They also use events with two exclusive jets, where jets
were constructed using an infrared unsafe jet
algorithm~\cite{Kunszt:1992tn}. In view of the considerable
uncertainties associated with the UA2 analysis we remain skeptical of
drawing negative conclusions. Instead we argue that our supersymmetric
D-brane construct could provide an explanation of the CDF anomaly if
acceptance and pseudorapidity cuts reduce the $Wjj$ production rate by
about 35\% and the UA2 90\% CL bound is taken as an order-of-magnitude
limit~\cite{UA2-comment}.

\begin{table}
\caption{Chiral couplings of $Y'$ and $Y''$ gauge bosons for $\phi = 0.0028$ and $g'_1 = 0.20$.}
\begin{tabular}{c|cc}
\hline
\hline
 Name &~~ $g_{Y'}Q_{Y'}$~ &~~$g_{Y''} Q_{Y''}$\\
\hline
~~$\bar U_i$~~ & $-0.368$ & $\phantom{-} 0.028$ \\[1mm]
~~$\bar D_i$~~ &  $-0.368 $ &  $\phantom{-} 0.209$ \\[1mm]
~~$L_i$~~ & $\phantom{-}0.143$ & $\phantom{-} 0.143$\\[1mm]
~~$\bar E_i$~~ & $-0.142$ & $- 0.262$\\[1mm]
~~$Q_i$~~ &  $\phantom{-} 0.368$ & $ -0.119$ \\[1mm]
\hline
\hline
\end{tabular}
\label{case3}
\end{table}

The $U(1)$ vector bosons couple to currents
\begin{eqnarray}
J_Y & = & 1.8 \times 10^{-1}~Q_{1R} + 5.9 \times 10^{-2}~Q_3 - 1.8 \times 10^{-1}~Q_{1L} \nonumber \\                    
J_{Y'}& = & 2.5 \times 10^{-4}~Q_{1R} + 3.7 \times 10^{-1}~Q_3 + 1.4\times 10^{-1}~Q_{1L}  \\ 
J_{Y''} & = & 9.0 \times 10^{-2}~Q_{1R} - 1.2 \times 10^{-1}~Q_3 + 3.5 \times 10^{-1}~Q_{1L} \nonumber  \, . 
\end{eqnarray}
Using Eq.~(\ref{bb-l}), we rewrite $J_{Y'}$ and $J_{Y''}$ as
\begin{eqnarray}
J_{Y'}& = & 2.5 \times 10^{-4}~Q_{1R} + 1.11~B + 1.4\times 10^{-1} L  \nonumber \\ 
J_{Y''} & = & 9.0 \times 10^{-2}~Q_{1R} - 2.5 \times 10^{-3}~(B+L) - 3.55 \times 10^{-1}~(B-L)  \, . 
\end{eqnarray}
Since ${\rm Tr}~[Q_{1R} \, B] = {\rm Tr}~[Q_{1R} L] = 0$, the decay widths are  given by
\begin{eqnarray}
\Gamma_{Y'} & = & \Gamma_{Y' \to Q_{1R}} + \Gamma_{Y' \to B} + \Gamma_{Y' \to L} \nonumber \\
& \propto & ( 2.5 \times 10^{-4})^2 \,\, {\rm Tr}~\left[Q_{1R}^2 \right] + (1.11)^2 \, \, {\rm Tr}~\left[B^2 \right] + (1.4\times 10^{-1})^2 \,\, {\rm Tr}~\left[L^2 \right] \nonumber \\
 & = & 8 \times ( 2.5 \times 10^{-4})^2   + \frac{4}{3} \times (1.11)^2 + 4 \times (1.4\times 10^{-1})^2 \,\nonumber  \\
& = & 5 \times 10^{-7} + 1.64 + 7.84 \times 10^{-2} \,,
\end{eqnarray}
\begin{eqnarray}
\Gamma_{Y''} & = & \Gamma_{Y'' \to Q_{1R}} + \Gamma_{Y' \to B +L} + \Gamma_{Y' \to B-L} \nonumber \\
& \propto & ( 9.0 \times 10^{-2})^2 \, {\rm Tr}[Q_{1R}^2] + (2.5 \times 10^{-3})^2  {\rm Tr}\left[(B +L)^2 \right] + (3.55\times 10^{-1})^2 {\rm Tr}\left[(B-L)^2 \right] \nonumber \\
& = & 6.48 \times 10^{-2} + 3.33 \times 10^{-5} + 6.7 \times 10^{-1} \, .
\end{eqnarray}
Thus, the corresponding branching fractions are
\begin{equation}
\begin{tabular}{lll} 
BR~$Y' \to Q_{1R}$  & : BR~$Y' \to B$ & : BR~$Y' \to L$ \\
$~~~2.9 \times 10^{-7}$ & : $~~~0.95$ & : $~~~0.046$
\end{tabular}
\end{equation}

and \begin{equation} 
\begin{tabular}{lll}
BR~$Y'' \to Q_{1R}$  & : BR~$Y'' \to B+L$ & : BR~$Y'' \to B-L$  \\
$~~~0.09$ & : $~~~4.5 \times 10^{-5}$ & : $~~~0.91$ \, .
\end{tabular}
\end{equation}
Of course, since the quiver construction has each particle straddling
two adjacent branes, there can be considerable variation in decay
channels particle by particle. The dominance of $B$ for the $Y'$ decay
channel and $B-L$ for the $Y''$ decay channel is valid after averaging
over decay channels.\footnote{An analogue is in the SM. The $Z$
  couples to a current $J_Z \propto T_3 - \tan^2 \theta_W \frac{Y}{2}$, where
  $Q = T_3 - \frac{Y}{2}$. In this case, $\sum (\frac{Y}{2})^2 = \frac{17}{6}$
  and ${\rm Tr} [T_3^2] = 2$; we have ${\rm BR}~Z \to T_3 : {\rm BR}~Z
  \to \frac{Y}{2} = 2 :\frac{17}{6} \tan^4 \theta_W = 2 : 0.25 =
  8:1$. However, this certainly does not hold particle by particle;
  e.g., for the neutrino electron doublet: $\Gamma_{Z \to \nu} \propto  (1 + \tan^2 \theta_W)^2 \sim 1.7$, whereas 
$\Gamma_{Z\to e} \propto (1 - \tan^2 \theta_W)^2 \sim 0.5$.} It is important to note that a
100\% coupling of the $Y'$ and $Y''$ to $B$ and $B-L$, respectively,
is possible only if the $U (1)$ gauge coupling constants are equal,
see Appendix~\ref{newappendix}.

\section{Leptophobic $\bm{Z'}$ at the LHC}
\label{LHC}

Since the CDF signal is in dispute, it is of interest to study the
predictions of the model for  energies not
obtainable at the Tevatron, but within the range of the LHC. To
illustrate the LHC phenomenology of our D-brane construct, we consider
the model in which the $U(1)$ right-handed neutrino has charges $Q_3 =
0,$ $Q_{1L}= Q_{1R} = -1$, i.e., $B-L$ is non-anomalous. 

The ATLAS Collaboration has searched for narrow resonances in the
invariant mass spectrum of dimuon and dielectron final states in event
samples corresponding to an integrated luminosity of $1.21~{\rm
  fb}^{-1}$ and $1.08~{\rm fb}^{-1}$,
respectively~\cite{Collaboration:2011dca}. The spectra are consistent
with SM expectations and thus upper limits on the cross section times
branching fraction for $Z'$ into lepton pairs have been set.

Using a data set with an integrated luminosity of $1~{\rm fb}^{-1}$,
the CMS Collaboration has searched for narrow resonances in the
dijet invariant mass spectrum~\cite{Chatrchyan:2011ns}. For $M_{Z'} \simeq
1~{\rm TeV}$, the CMS experiment has excluded production rates
$\sigma (p p) \times {\rm BR} (Z' \to jj) \times {\cal A} > 1~{\rm
  pb}$ at the 95\%CL.  Each event in the search is required to have
its two highest-$p_T$ jets with (pseudorapidity) $|\eta_j| < 2.5$ and
the leading jet must satisfy $p_t^{j_1} > 150~{\rm GeV}$, with
$|\Delta \eta_{jj}| < 1.3$.  The acceptance ${\cal A}$ of selection
requirements is reported to be $\approx 0.6$.

\begin{figure}[tbp] \begin{minipage}[t]{0.49\textwidth} \postscript{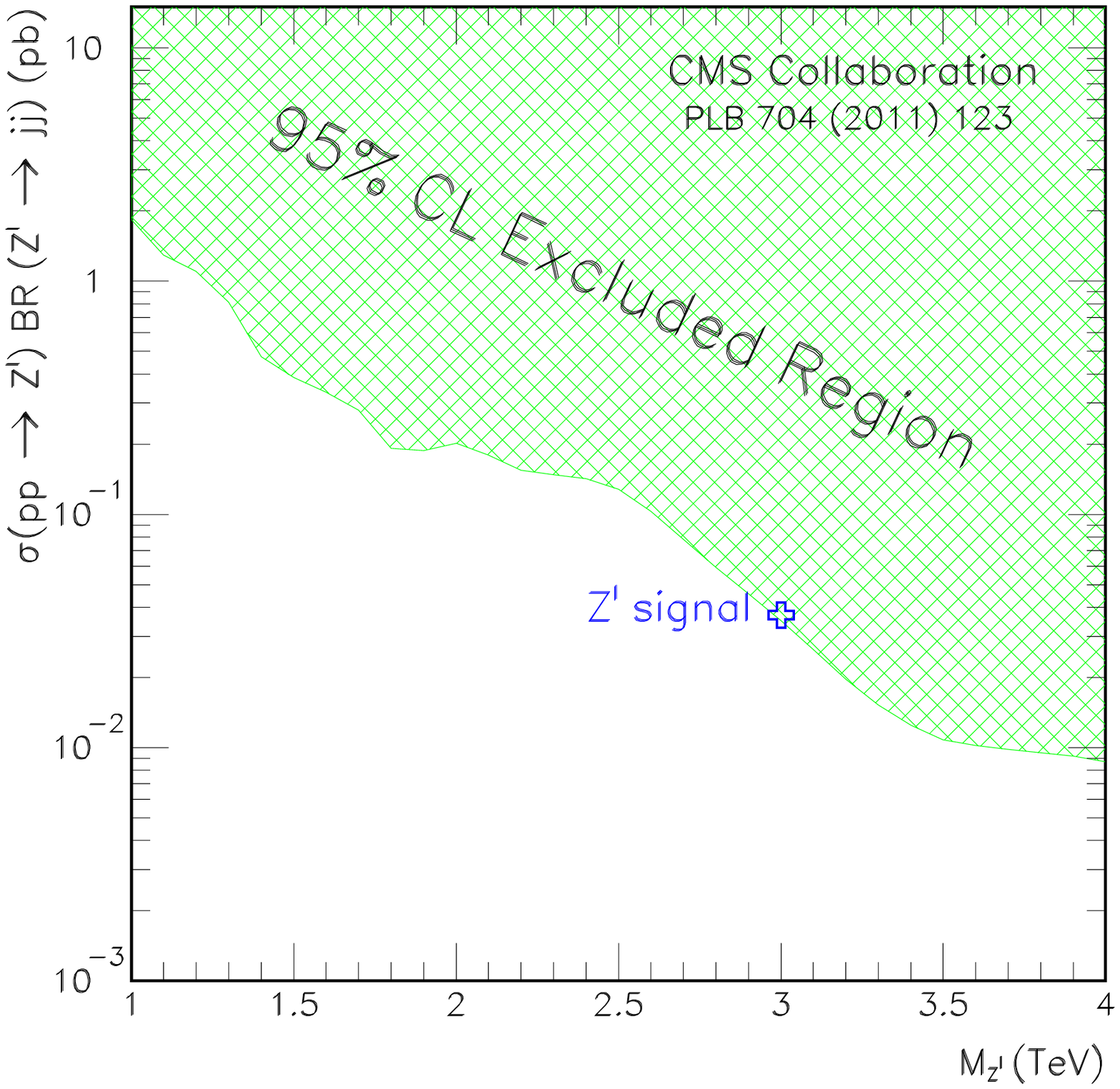}{0.99} \end{minipage} \hfill \begin{minipage}[t]{0.49\textwidth} \postscript{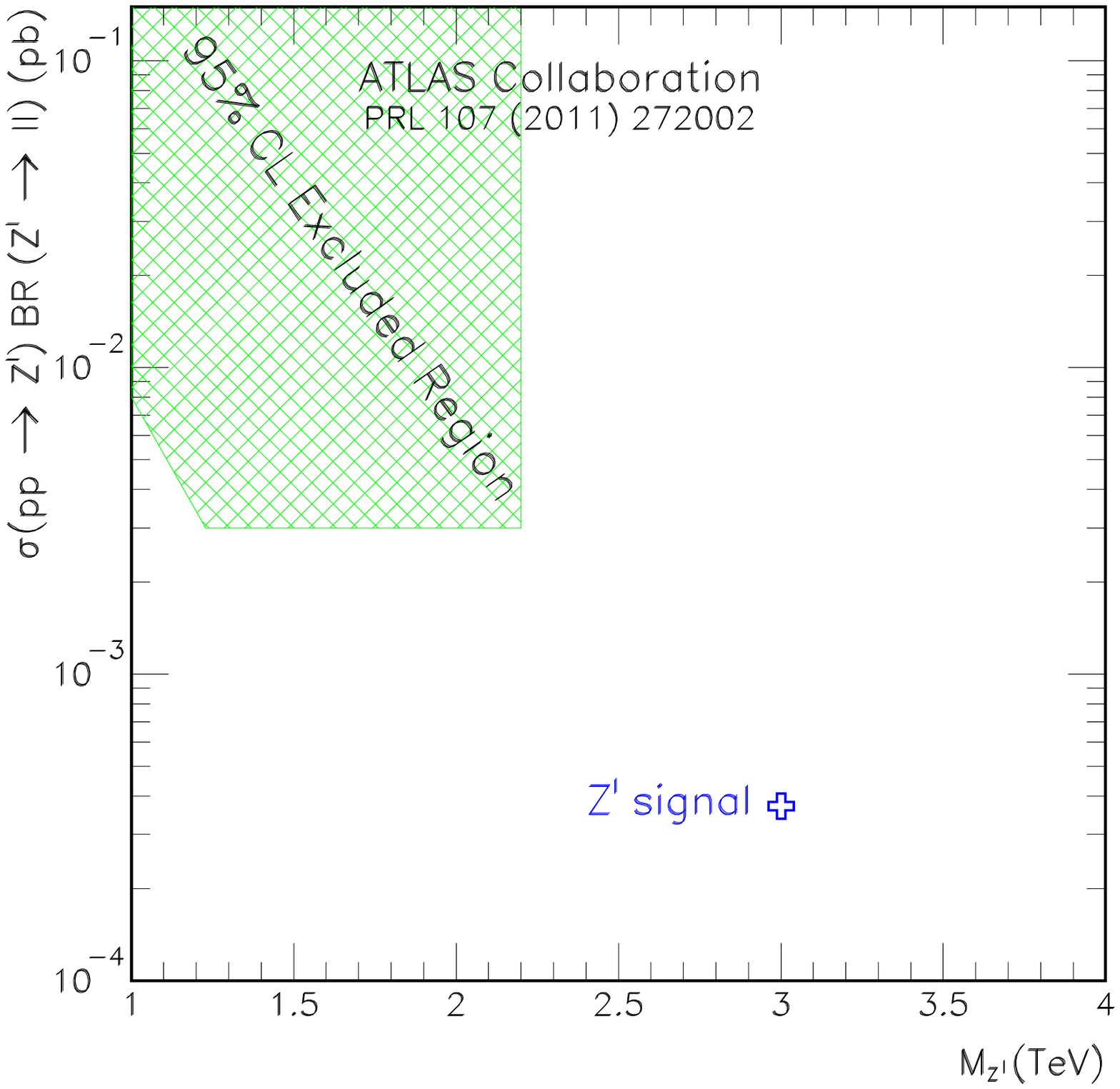}{0.99} \end{minipage} \caption{Comparison of the (pre-cut) total cross section for the production of $p p \to Z' \to jj$ (left) and $p p \to Z' \to \ell \ell$ (right) with the 95\% CL upper limits on the production of a gauge boson decaying into two jets (left) and two leptons (right), as reported by the CMS (corrected by acceptance)~\cite{Chatrchyan:2011ns} and ATLAS~\cite{Collaboration:2011dca} collaborations, respectively. We have taken $\phi = 0.0028$, $g'_1 = 0.20$. For isotropic decays (independently of the resonance), the acceptance for the CMS detector has been reporetd to be ${\cal A} \approx 0.6$~\cite{Chatrchyan:2011ns}. The predicted $Z'$ production rates for $\sqrt{s} = 7~{\rm TeV}$
and $M_{Z'} \simeq 3~{\rm TeV}$ saturate the current limits.}  
\label{LHC-Zprime} \end{figure}

To compare our predictions  with LHC experimental
searches in dilepton and dijets it is sufficient to consider the
production cross section in the narrow $Z'$ width approximation,
\begin{equation}
\hat \sigma (q \bar q \to Z')  =   K \frac{2 \pi}{3} \, \frac{G_F \, M_Z^2}{\sqrt{2}}  \left[v_q^2 (\phi, g'_1)+ a_q^2 (\phi, g'_1) \right] \, \delta \left(\hat s - M_{Z'}^2 \right) \,,
\end{equation}
where $G_F$ is the Fermi coupling constant and the $K$-factor represents the enhancement from higher order QCD
processes estimated to be $K \simeq 1.3$~\cite{Barger}. After folding
$\hat \sigma$ with the CTEQ6 parton distribution
functions~\cite{Pumplin:2002vw}, we determine (at the parton level)
the resonant production cross section. In Fig.~\ref{LHC-Zprime} we
compare the predicted $\sigma (p\bar p \to Z') \times {\rm BR} (Z' \to
jj)$ (left panel) and $\sigma (p\bar p \to Z') \times {\rm BR} (Z' \to
\ell \ell)$ (right panel) production rates with 95\% CL upper limits
recently reported by the CMS~\cite{Chatrchyan:2011ns} and
ATLAS~\cite{Collaboration:2011dca} collaborations. Selection cuts will
probably reduced event rates by factors of 20\%. Keeping this in mind,
we conclude that the 2012 LHC7 run will probe $ M_{Z'} \sim 3~{\rm
  TeV}$, whereas future runs from LHC14 will provide a generous
discovery potential of up to about $M_{Z'} \sim 8~{\rm TeV}.$

\section{Regge excitations}
\label{SRegge}

\begin{figure}[tpb]
 \postscript{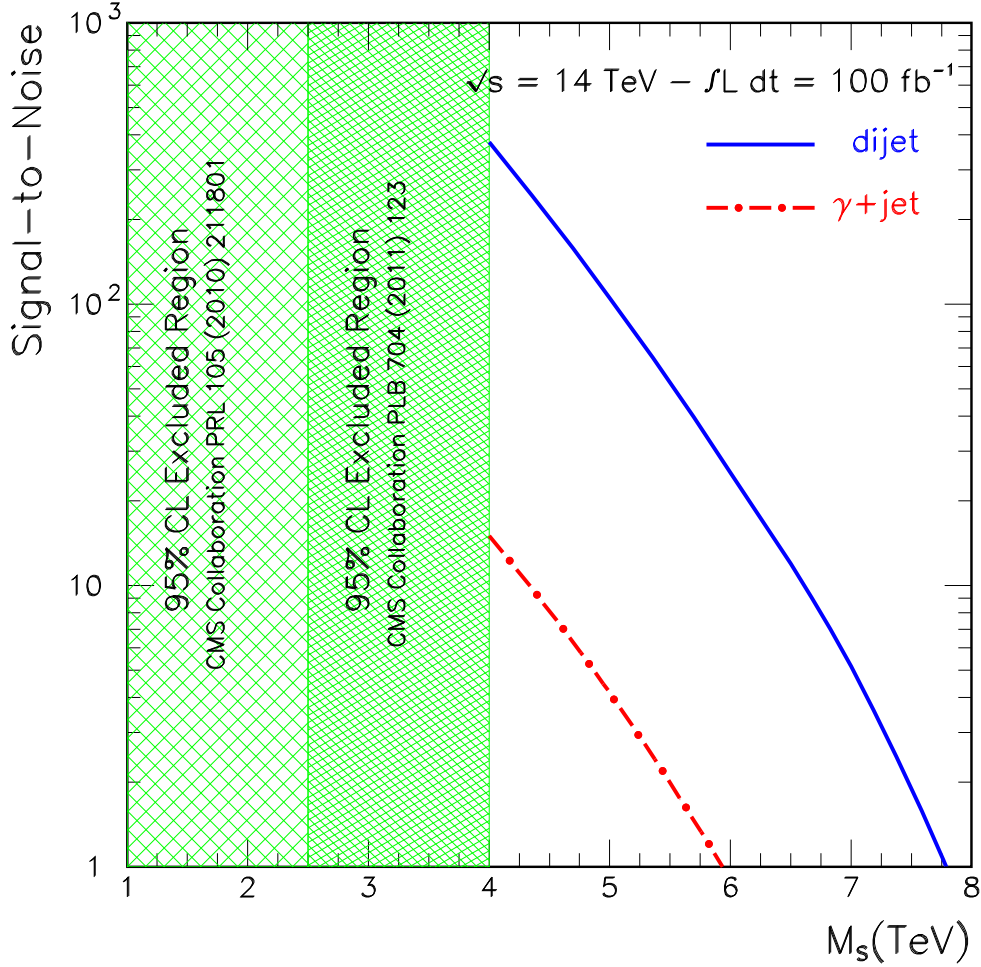}{0.8}
\caption{Signal-to-noise ratio of $pp \to {\rm
    dijet}$ and $pp \to \gamma + {\rm jet}$, for $\sqrt{s} = 14$~TeV, ${\cal L} = 100~{\rm fb}^{-1}$, and  $\kappa^2
  \simeq 0.02$. The approximate
  equality of the background due to misidentified $\pi^0$'s and the
  QCD background, across a range of large $p_T^\gamma$ as implemented
  in~\cite{Anchordoqui:2007da}, is maintained as an approximate
  equality over a range of  $\gamma$-jet invariant masses
  with the rapidity cuts imposed ($|y_{\rm max}^{j}| < 1.0$ and $|y_{\rm max}^{\gamma}| < 2.4$). Details of the  signal and background calculations have been given elsewhere~\cite{Anchordoqui:2008di}.}
\label{Regge}
\end{figure}

In TeV-scale gravity scenarios where the SM is realized on the
world-volume of D-branes the presence of fundamental strings can also
be unearthed by searching for the effects of their vibrations. The
particles that appear as the quanta of oscillating string modes are
called Regge excitations and have squared masses quantized in units of
$M_s = 1 /\sqrt{\alpha'}$, where $\alpha'$ is the Regge slope
parameter~\cite{Veneziano:1968yb}. The leading contributions of Regge
recurrences to certain processes at hadron colliders are {\em
  universal}. This is because the full-fledged string amplitudes which
describe $2 \to 2$ parton scattering subprocesses involving four gauge
bosons as well as those with two gauge bosons and two chiral matter
fields are (to leading order in string coupling, but all orders in
$\alpha'$) independent of the D-brane configuration, the geometry of
the extra dimensions, and whether supersymmetry is broken or
not. Therefore, the $s$-channel pole terms of the average square
amplitudes contributing to dijet production can be obtained
independent of the details of the compactification
scheme~\cite{Lust:2008qc}. For phenomenological purposes, the poles
need to be softened to a Breit-Wigner form by obtaining and utilizing
the corrcet {\it total} widths of the
resonances~\cite{Anchordoqui:2008hi}. After this is done, it is
feasible to compute genuine string corrections to dijet signals at the
LHC~\cite{Anchordoqui:2008di}.\footnote{Phenomenological studies of
  Regge excitations and associated collider signatures, based on
  simple toy model embedding parts of the SM into string theory, have
  been carried out in~\cite{Cullen:2000ef}. The discovery potential of
  string resonances via top quark pair production in the context of
  cannonical D-brane constructions has been recently
  established~\cite{Dong:2010jt}.}  The CMS Collaboration has searched
for such narrow resonances in their dijet mass
spectrum~\cite{Khachatryan:2010jd}. After operating for only few
months, with merely 2.9 inverse picobarns of integrated luminosity,
the LHC CMS experiment has ruled out $M_s < 2.5~{\rm TeV}$. The LHC7
has recently delivered an integrated luminosity in excess of $1~{\rm
  fb}^{-1}$. This extends considerably the search territory for new
physics in events containing dijets. The new data exclude string
resonances with $M_s < 4~{\rm TeV}$~\cite{Chatrchyan:2011ns}. In fact,
as shown in Fig.~\ref{Regge}, the LHC has the capacity of discovering
strongly interacting resonances via dijet final states in practically
all range up to $\frac{1}{2}\sqrt{s}_{\rm LHC}$. Of particular
interest here, for the $U(3)_C \times Sp(1)_L \times U(1)_L \times
U(1)_R$ D-brane model, the anomaly cancelation fixes the projection of
the hypercharge into the color stack at the string scale: $\kappa =
c_3 \sqrt{6} g_Y/g_3$~\cite{Anchordoqui:2010zs}.  Therefore one can
also cleany extract the leading string corrections to $\gamma$ + jet
signals at the LHC~\cite{Anchordoqui:2007da}.\footnote{ {\it E.g.},
  the tree level amplitude for gluon fusion into $\gamma +$ jet, $
  {\cal M} (gg \to \gamma g) = \cos \theta_W \, \, {\cal M} (gg \to Y
  g) = \kappa \, \, \cos \theta_W \, \, {\cal M} (gg \to Cg),$ has a
  unique free parameter that is the string scale $M_s$. Here,
  $\theta_W$ is the weak angle and $C$ is the extra $U(1)$ boson tied
  to the color stack. For details see~\cite{Anchordoqui:2007da}.}  The
precise predictions for the branching fraction of two different
topologies (dijet and $\gamma +$ jet) can be used as a powerful
discriminator of low mass string excitations from other beyond SM
scenarios.

\section{Conclusions}
\label{conclusions}

We have shown that a $Z'$ that can explain the CDF $Wjj$
excess and is in full agreement with existing limits on $Z'$
couplings to quarks and leptons can materialize in the context of
D-brane TeV-scale string compactifications. The existence of 
additional, largely leptophopic, $Z$'s with anomalous masses
somewhat less than the string scale, is generic to the D-brane
models discussed in some detail in this paper. Thus, even if the CDF
anomaly does not survive additional scrutiny, there may exist such
$Z$'s with masses $\agt~1~{\rm TeV}$, whose discovery is out of reach of
Tevatron, but open to such at LHC. In that case the analysis
presented here can be directly applied to the higher energy realm,
with a view toward identifying the precise makeup of the various
abelian sectors, and pursuing with strong confidence a signal at LHC
for the Regge excitations of the string. We have long imagined
strings to be minuscule objects which could only be experimentally
observed in the far-distant future. It is conceivable that this
future has already arrived.

\section*{Acknowledgments}
We would like to thank Fernando Quevedo and Timo Weigand for useful
discussions. L.A.A.\ is supported by the U.S.  National Science
Foundation (NSF) under Grant PHY-0757598 and CAREER Award
PHY-1053663. I.A.\ is supported in part by the European Commission
under the ERC Advanced Grant 226371 and the contract PITN-GA-2009-
237920. H.G.\ and T.R.T.\ are supported by NSF Grant PHY-0757959.
X.H.\ is supported in part by the National Research Foundation of
Korea grants 2005-009-3843, 2009-008-0372, and 2010-220-C00003. D.L.\
is partially supported by the Cluster of Excellence "Origin and
Structure of the Universe", in Munich. D.L.\ and T.R.T.\ thank the
Theory Department of CERN for its hospitality.  Any opinions,
findings, and conclusions or recommendations expressed in this
material are those of the authors and do not necessarily reflect the
views of the National Science Foundation.

\appendix

\section{Properties of the Anomalous Mass Sector}

Outside of the Higgs couplings, the relevant parts of the Lagrangian are the gauge couplings
generated by the $U(1)$ covariant derivatives acting on the matter fields, and the
(mass)$^2$ matrix of the anomalous sector
\begin{equation}
{\cal L} = Q^T\  g\ X  + \tfrac{1}{2} X^T \mathbb{M}^2 X \,,
\end{equation}
where $X_i$ are the three $U(1)$ gauge fields in the D-brane basis ($B_\mu,\, C_\mu,\, \tilde B_\mu$),
$g$ is a diagonal coupling matrix $(g'_1, g'_3, g'_4)$, and $Q$ are the 3 charge matrices.

As in Sec.~\ref{SII}, perform a rotation $X= {\cal R} Y$ and require that one of the $Y$'s (say $Y_\mu$) couple to hypercharge. We then obtain the constraint on the first column of ${\cal R}$ given in Eq.~(\ref{ocho}). However, there is now an additional constraint: {\em the field $Y_\mu$ is an eigenstate of $\mathbb{M}^2$ with zero eigenvalue.} Under the ${\cal R}$ rotation, the mass term becomes \begin{equation} \tfrac{1}{2} X^T \mathbb{M}^2 X = \tfrac{1}{2} Y^T\ \overline{\mathbb{M}^2}\ \ Y \,, \end{equation} with $\overline{\mathbb{M}^2} = {\cal R}^T\ \mathbb{M}^2 \ {\cal R} .$ We know that at least $Y_\mu$ is an eigenstate with eigenvalue $0$. We also know that Poincare invariance requires the complete diagonalization of the mass matrix in order to deal with observables. However, further similarity transformations will undo the coupling of the zero eigenstate to hypercharge. There seems no way of eventually fulfilling all these conditions except to require that the same ${\cal R}$ which rotates to couple $Y_\mu$ to hypercharge simultaneously diagonalizes $\mathbb{M}^2$ so that \begin{equation} \overline{\mathbb{M}^2} = {\rm diag} (0, M'^2, M''^2) \, .  \end{equation} This implies that the original $\mathbb{M}^2$ in the D-brane basis is given by \begin{equation} \mathbb{M}^2 = {\cal R}\ {\rm diag} (0, M'^2, M''^2) {\cal R}^T \,, \end{equation} which results in the following baroque matrix: \begin{equation} \mathbb{M}^2 = \left( \begin{array}{ccc} a & b & c \\ b& d & e \\ c & e & f \end{array} \right) \,, \end{equation} where \begin{eqnarray}
  a & = &  M'^2 (C_\psi S_\theta S_\phi - C_\phi S_\psi)^2 + M''^2 (C_\phi C_\psi S_\theta + S_\phi S_\psi)^2 \,, \nonumber \\
 b & = & ({M'}^2-{M''}^2) C_\phi C_{2\psi} S_\theta S_\phi + C^2_\phi C_\psi (-{M'}^2 +{M''}^2 S^2_\theta) S_\psi + C_\psi 
(-{M''}^2 + {M'}^2 S_\theta^2) S_\phi^2 S_\psi  \, , \nonumber \\
c & = &  C_\theta [{M''}^2 C^2_\phi C_\psi S_\theta + {M'}^2 C_\psi S_\theta S^2_\phi - ({M'}^2-{M''}^2) C_\phi S_\phi S_\psi] \,, \nonumber\\
d & = & {M''}^2 (C_\psi S_\phi - C_\phi S_\theta S_\psi)^2 + {M'}^2 (C_\phi C_\psi + S_\theta S_\phi S_\psi)^2 \, , \nonumber \\
e & = & C_\theta [({M'}^2 - {M''}^2) C_\phi C_\psi S_\phi + {M''}^2 C_\phi^2 S_\theta S_\psi + {M'}^2 S_\theta S_\phi^2 S_\psi ] \, , \nonumber \\
 f & = & C_\theta^2 ({M''}^2 C^2_\phi + {M'}^2 S^2_\phi ) \, .
\label{biguglyMF}
\end{eqnarray}

\section{$\bm{B}$ and $\bm{B-L}$ couplings on the rotated basis}
\label{newappendix}

For given a set of $U(1)$ fields with orthogonal charges in the $1,\,
2,\, 3, \dots$ basis, an obvious question is whether each of the
fields on the rotated basis couples to a single charge $\bar Q_i$. Let
\begin{equation}
{\cal L}  = X^T \, g \, Q \,,
\label{milito}
\end{equation}
be the Lagrangian in the $1,\ 2,\ 3,\ \dots$ basis, with $X_\mu^i$ and  
$Q_i$ vectors and $g$ a diagonal matrix in $N$-dimensional 'flavor' space.  
Now rotate to new orthogonal basis ($\bar{Q}$) for $Q$
\begin{equation}
Q = {\cal O} \, \bar{Q} \,;
\end{equation}
(\ref{milito}) becomes
\begin{equation}
{\cal L}  = X^T \, g \, {\cal O} \, \bar{Q} \, .
\end{equation}
As it stands, each $X_\mu^i$ does not couple to a unique charge $\bar Q_i$; hence we rotate $X$, 
\begin{equation}
X = {\cal R} \, \bar{Y},
\end{equation}
to obtain
\begin{equation}
{\cal L} = \bar{Y}^T \, {\cal R}^T \, g \, {\cal O} \, \bar{Q} \, .
\end{equation}
We wish to see if, for given ${\cal R}$ and $g$, we can find an ${\cal
  O}$ so that
\begin{equation}
{\cal R}^T \, g \, {\cal O} = \bar{g} \, ({\rm diagonal}) \, .
\label{piletaya}
\end{equation}
This allows each $\bar Y_\mu^i$ to couple to a unique charge $\bar Q_i$ 
with strength $\bar g_i$. To see the problem with this, we rewrite 
(\ref{piletaya}) in terms of components
\begin{equation}
({\cal R}^T)_{ij} \,\, g_j \,\, {\cal O}_{jk} = \bar g_i \, \delta_{ik} \, ;
\label{MORSANADA}
\end{equation}
for $i \neq k$, (\ref{MORSANADA}) leads to
\begin{equation}
({\cal R}^T)_{ij} \,\, g_j\, {\cal O}_{jk} = 0 \ .
\label{smores}
\end{equation}
In general, in Eq.~(\ref{smores}) there are $N(N-1)$ equations, but only $N (N-1)/2$ independent ${\cal R}_{ij}$ generators in $SO(N)$; therefore the system is overdetermined. Of course, if $g = g \mathds{1}$, the equation becomes
\begin{equation}
{\cal R}^T \, {\cal O} = \mathds{1},
\end{equation}
and so ${\cal O} = {\cal R}$.

\end{document}